\documentclass[twocolumn,showkeys,preprintnumbers,floatfix,amsmath,a4paper,nofootinbib,amssymb]{revtex4}

\usepackage[latin1]{inputenc}
\usepackage{graphicx}% Include figure files
\usepackage{bm}% bold math
\usepackage{epstopdf}

\begin{document}

\title{Understanding scaling through history-dependent processes with collapsing sample space}
\author{Bernat Corominas-Murtra$^1$, Rudolf Hanel$^1$ and Stefan Thurner$^{1,2,3}$}
%\author{Authors?}

\thanks{stefan.thurner@meduniwien.ac.at}

\affiliation{
$^1$ Section for Science of Complex Systems; Medical University of Vienna, Spitalgasse 23; A-1090, Austria\\
$^2$ Santa Fe Institute; 1399 Hyde Park Road; Santa Fe; NM 87501; USA. \\
$^3$IIASA, Schlossplatz 1, A-2361 Laxenburg; Austria.}

%%%%%%%%%%%%%%%%%%%%%%%%%%%%%%%%%%%%%%%%%%%%%%%%%%%%%%%%%%%%%%%%

\begin{abstract}
History-dependent processes are ubiquitous in natural and social systems. Many such stochastic processes, especially those that are associated with 
complex systems, become more constrained as they unfold, meaning that their sample-space, or their set of possible outcomes, reduces as they age. 
We demonstrate that these sample-space reducing (SSR) processes necessarily lead to Zipf's law in the rank distributions of their outcomes. 
We show that by adding noise to SSR processes the corresponding rank distributions remain exact power-laws, $p(x)\sim x^{-\lambda}$, 
where the exponent directly corresponds to the mixing ratio of the SSR process and noise. 
This allows us to give a precise meaning to the scaling exponent in terms of the degree to how much a given process reduces 
its sample-space as it unfolds. Noisy SSR processes further allow us to explain a wide range of scaling exponents in frequency distributions 
ranging from $\alpha = 2$ to $\infty$. 
We discuss several applications showing how SSR processes can be used to understand Zipf's law in 
word frequencies, and how they are related to diffusion processes in directed networks, or ageing processes such as in fragmentation processes. 
SSR processes provide a new alternative to understand the origin of scaling in complex systems without the 
recourse to multiplicative, preferential, or self-organised critical processes. 
\end{abstract}
%\pacs{05.10.-a, 05.40.-a, 05.50.+q, 05.65.+b}

%
%\subsection{Significance statement}
%Many complex systems reduce their flexibility over time in the sense that the number of options (possible states)
%diminishes over time. We show that rank distributions of the visits to these states that emerge from such processes 
%are exact power-laws with an exponent -1, which is called Zipf's law. When noise is added to such processes, 
%meaning that from time to time they can also increase the number of their options, the rank distribution remains a power-law, 
%with an exponent that is related to the noise level in a remarkably simple way.  Sample-space reducing 
%processes provide a new route to understand the phenomenon of 
%scaling, and provides an alternative to the known mechanisms of self-organized criticality, multiplicative processes, or 
%preferential attachment.
\maketitle

A typical feature of ageing is that the number of possible states in a system reduces as it ages.  
While a newborn can become a composer, politician, physicist, actor, or anything else, 
the chances for a 65 year old physics professor to become a concert pianist are practically zero.   
A characteristic feature of history-dependent systems is that their sample-space, defined as 
the set of all possible outcomes, changes over time. 
Many ageing stochastic 
systems (such as career paths), become more constrained in their dynamics as they unfold, i.e., their sample-space becomes smaller over time. 
An example for a sample-space reducing process is the formation of sentences. The first word in a sentence 
can be sampled from the sample-space of all existing words. The choice of subsequent words is constrained by grammar and context, 
so that the second word can only be sampled from a smaller sample-space. 
As the length of a sentence increases, the size of sample-space of word use typically reduces.  
\begin{figure}[ht!]
\begin{center}
\includegraphics[width= 7.7cm]{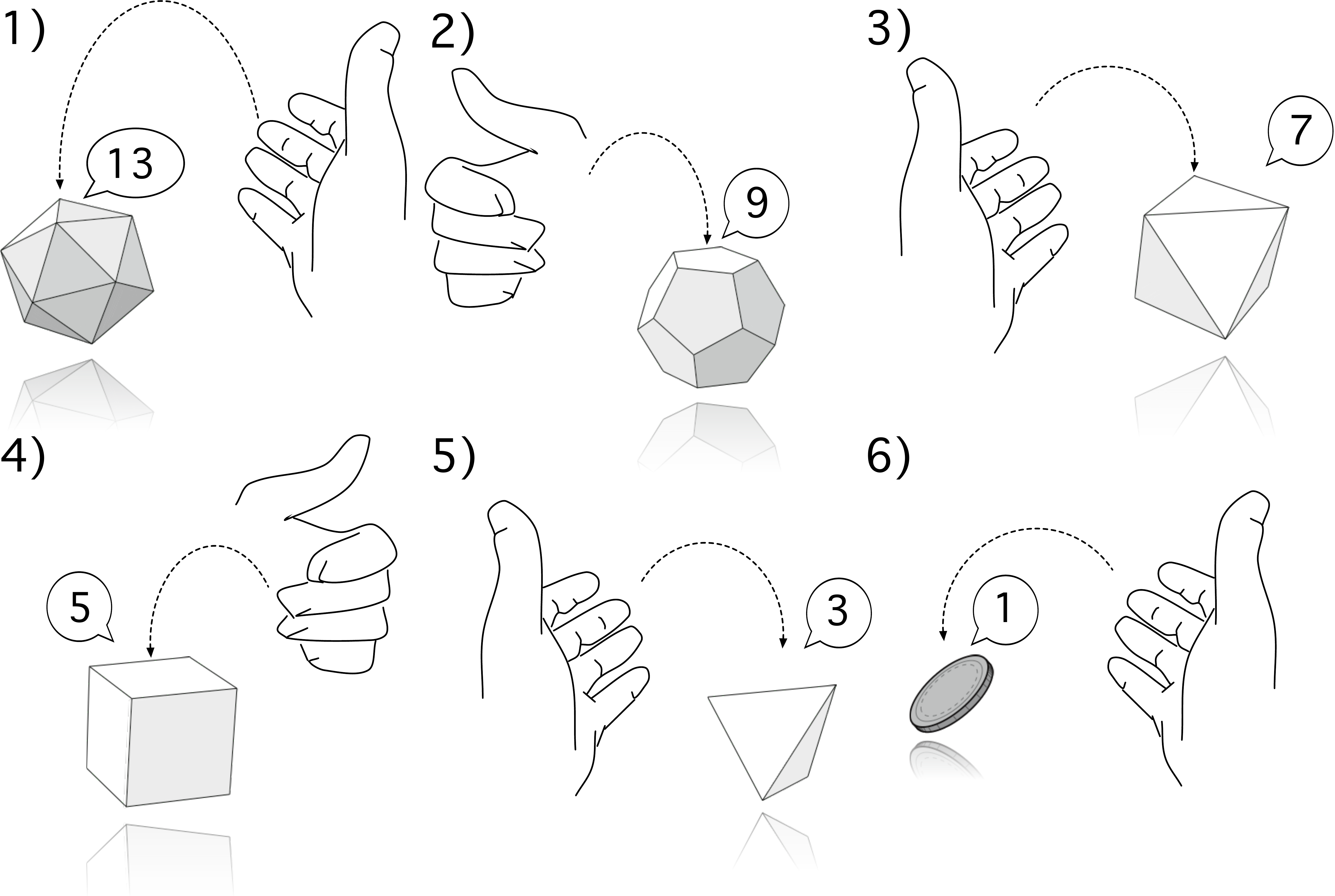}
\caption{
Sample-space reducing process. Imagine a set of $N=20$ dice with different numbers of faces. We start by throwing the 20-faced dice (icosahedron). 
Suppose we get a face-value of 13. We now have to take the 12-faced dice (dodecahedron), throw it, and get a face-value of 
say 9, so that we must continue with the 8-faced dice. Say we throw a 7, forcing us to take the (ordinary) dice, with which we 
throw say a 5. With the 4-faced dice we get a 2, which forces us to take the 2-faced dice (coin). The process ends when we 
throw a 1 for the first time. The set of possible outcomes (sample-space) reduces as the process unfolds. The sequence above 
was chosen to make use of the platonic dice for pictorial reasons only. If the process is repeated many times, 
the distribution of face-values (rank ordered) gives Zipf's law.}
\label{Fig:dice}
\end{center}
\end{figure}

Many  history-dependent processes are characterised by power-law distribution functions in their frequency and 
rank distributions of their outcomes. The most famous example is the rank distribution of word frequencies in  
texts, which follows a power-law with an approximate exponent of $-1$, the so-called Zipf's law \cite{Zipf:1949}. Zipf's law 
has been found in countless natural and social phenomena, including gene expression patterns \cite{stanley1999},  
human behavioural sequences \cite{thurner2012}, fluctuations in financial markets  \cite{Gabaix:2003}, 
scientific citations \cite{price,redner},  distributions of city- \cite{Makse:1995}, and firm sizes \cite{Axtell:2001,saichev2008}, 
and many more, see e.g. \cite{Newman:2005}\footnote{Some of these examples are of course not  associated with sample-space reducing processes.}.
Over the past decades there has been a tremendous effort to understand the origin of power-laws in distribution functions 
obtained from complex systems. Most of the existing explanations are based on multiplicative processes \cite{simon:1955,mandelbrot1953,Levy:1996,pietronero}, 
preferential mechanisms \cite{Biham:1999,solar,barabasi}, or self-organised criticality \cite{Bak:1987,Corominas-Murtra:2010, Corominas-Murtra:2011}.
Here we offer an alternative route to understand scaling based on processes that reduce their sample-space over time. 
We show that the emergence of power-laws in this way is related to the breaking of a symmetry in random sampling processes, a mechanism 
that was explored  in \cite{Hanel:2014}. History-dependent random processes have been studied generically \cite{kac,clifford}, however not 
with the  rationale to understand the emergence of scaling in complex systems. 
%

%%%%%%%%%%%%%%%%%%%%%%%%%%%
\begin{figure}[t]
	\begin{center}
		\includegraphics[width= 8.5cm]{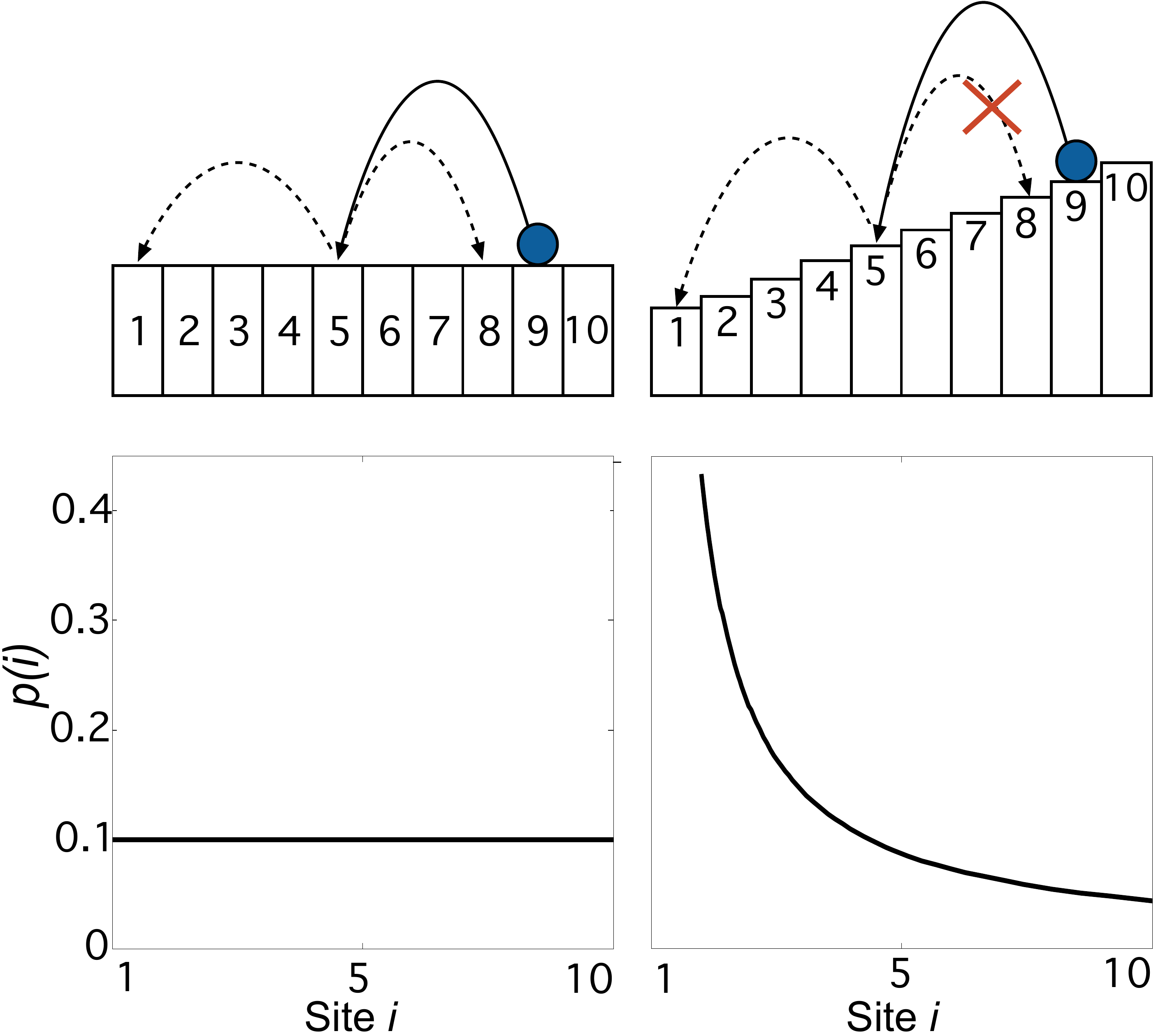}
		\caption{Illustration of path dependence, sample-space reduction, and nestedness of sample-space. 
			(Left)  Unconstrained (iid) random walk $\phi_R$ realized by a ball randomly bouncing between all 
			possible sites. The probability to observe the ball at a given site $i$ is uniform, $p(i)=1/N$. 
			(Right) The ball can only bounce downward, the left-right symmetry is broken. When  
			level $1$ is reached the process stops and is repeated. Sample-space reduces from step to step in a nested way 
			(main feature of SSR processes). After many iterations the occupation distribution (visits to level $i$) follows Zipf's 
			law, $p_{N=10}(i)\propto i^{-1}$. Symmetry breaking of the sampling changes the uniform probability distribution to a power-law. 
		} 
		\end{center}
			\label{Fig:MinimalSyst}
\end{figure}
%%%%%%%%%%%%%%%%%%%%%%%%%%%

\section{Results}

The essence of SSR processes can be illustrated by a set of $N$ fair dice with 
different numbers of faces. The first dice has one face, the second has two faces (coin), the third one three, 
etc., up to dice number $N$, which has $N$ faces. The faces of a dice are numbered and have respective face values. 
To start the SSR process, take the dice with the largest number of faces ($N$) and throw it. The result is a face value between $1$ and $N$, 
say it is $K$. We now take dice number $K-1$ (with $K-1$ faces) and throw it, to get a number $i$ between $1$ and $K-1$, 
say we throw $L$. We now take dice number $L-1$ throw it, etc. We repeat the process until we reach dice number $1$, 
and the process stops. We denote this directed and acyclic process by $\phi$. As the process unfolds, $\phi$ generates a 
single sequence of 
strictly decreasing numbers $i$. An intuitive realisation of this process is depicted in Fig. 1. 
The probability that the process $\phi$ visits the particular site $i$ in a sequence is the {\em visiting} probability $P_N(i)$, which can  
easily be shown to follow an exact Zipf's law, $P_N(i)=1/i$. 
This is shown with a simple proof by induction on $N$. Take the process $\phi$ and let $N=2$. 
There exist two possible sequences: Either $\phi$  directly generates a $1$ with a probability 
$1/2$, or $\phi$ first generates $2$ with probability $1/2$, and then a $1$ with certainty.
Both sequences visit $1$ but only one visits $2$. As a consequence, $P_2(2)=1/2$ and $P_2(1)=1$.
Let us now suppose that $P_{N'}(i)=1/i$ has been shown up to level $N'=N-1$.
Now, if the process starts with dice $N$, the probability to hit $i$ in the first step is $1/N$. 
Also, any other $j$, $N \geq j>i$, is reached with probability $1/N$. 
If we get  $j>i$, we will obtain $i$ in the next step with probability $P_{j-1}(i)$, 
which leads us to the recursive scheme for all $i<N$, 
$P_{N}(i)=\frac{1}{N}\left(1+\sum_{i<j\leq N}P_{j-1}(i)\right).$
Since by assumption $P_{j-1}(i)=1/i$, with $i< j\leq N$ holds, simple algebra yields $P_{N}(i)=1/i$. 
Finally, as pointed out above, for $i=N$, we have $P_{N}(N)=1/N$, which completes the proof that 
indeed the visiting probability is
\begin{equation}
P_{N}(i)=\frac{1}{i} \quad .
\end{equation}

If the process $\phi$ is repeated many times, meaning that once it reaches dice number $1$, we start by throwing dice number $N$ again, 
we are interested in how often a given site $i$ is occupied on average. The  {\em occupation} probability for site $i$, given that there are 
$N$ possible sites, is denoted by $p_N(i)$. 
Note an important property of the process $\phi$. While in general the visiting probability $P_N$ and the occupation probability $p_N$
of a process quantify different aspects, for the particular process $\phi$ both probabilities only differ by a normalization factor. 
This is so because any sequence generated by $\phi$ is strictly decreasing and contains any particular site $i$ at most once. 
Further, any sequence ends on site $1$, meaning $P_N(1)=1$. Therefore, it is clear that $P_N(i)=p_N(i)/p_N(1)$, where $p_N(1)$
is a normalisation factor. 
This shows that this prototype of a SSR processes exhibits an exact Zipf's law in the (rank ordered) occupation probabilities. 

An alternative picture that illustrates the history-dependence aspect of the same SSR 
processes is shown in Fig. 2. 
In the left panel we show an iid stochastic process, where the space of potential outcomes is $\Omega=\{1, . . .,N\}$. 
At each timestep a ball can jump from one of $N$ sites of $\Omega$ to any other with equal probability. Since the process is 
independent, the conditional probability of jumping from site $i$ to site $j$ is  ${P}(j|i)=1/N$. There is no path dependence. 
If we define $\Omega_i$ as the subset of those sites that can be reached from site $i$, we obviously find that this is 
constant over time,
\[  	\Omega_1=\Omega_2= . . .=\Omega_N =\Omega \quad . \]
We refer to this process as an {\em unconstrained random walk} and denote it by $\phi_R$. 
The visiting distribution is $p(i)=1/N$, see Fig. 2.  
To introduce path- or history dependence, assume that sites are arranged in levels like a staircase.
Now imagine a ball that can bounce {\em downstairs} to lower levels randomly, 
but never can climb to higher levels,  Fig. 2 (right panel). 
If at time $t$ the ball is at level (site) $i$, at $t+1$ all lower levels $j <i$ can be 
reached with the same probability, ${P}(j|i)=1/(i-1)$. 
Jumps to higher levels are forbidden, ${P}(j|i)=0$, for $j\geq i$. 
The process ends at the lowest stair level 1. 
In this process, sample-space displays a nested structure,  
\[ 
	\Omega_1\subset \Omega_2\subset . . . \subset\Omega_{N} \subset \Omega \quad.
	\label{eq:nest}
\]
In this case, $\Omega_{i}=\{1,2,\cdots,i-1\}$, for all values of $i\in\Omega$. 
$\Omega_1$  is the empty set. 
This nested structure of sample-space is the defining property of SSR processes. 
This type of nesting breaks the left-right symmetry of the iid stochastic process. 
The visiting probability to sites (levels) $i$ during a downward sequence is 
again $P_N(i)=1/i$. Since this process is equivalent to $\phi$, the same proof applies. 

It is conceivable that in many real systems nestedness of SSR processes is not realized perfectly  
and that from time to time the sample-space can also expand during a sequence. 
In the above example this would mean that from time to time random up-ward moves are allowed, 
or equivalently, that the nested process $\phi$ is perturbed by noise. 
In the context of the scenario depicted in Fig. 2 we look at a superposition of 
the SSR $\phi$, and the unconstrained random walk $\phi_R$. 
Using $\lambda$ to denote the mixing ratio, the nested process $\Phi^{(\lambda)}$ with noise is written as
\begin{equation}
	\Phi^{(\lambda)}=\lambda\phi+(1-\lambda)\phi_R \quad , \quad \lambda\in[0,1] \quad .
	\label{vieh}
\end{equation}
More concretely, if the ball is at site $i$, with probability $\lambda$ it jumps (downward) to any of site $k \in \Omega_{i}$ 
(with uniform probability), and with probability $1-\lambda$, it jumps to any of the $N$ sites, ($j \in\Omega)$. 
In other words, each time before throwing the dice we decide with probability $\lambda$ that the sample-space 
for the next throw is $\Omega_{i}$ (SSR process), or with $(1-\lambda)$ it is $\Omega$ (iid noise $\phi_R$). 
We repeat this process until the face value $1$ is obtained. 
With probability $\lambda$ the process is $\phi$ and stops, 
and with probability $(1-\lambda)$ the process is $\phi_R$ and continues until $1$ occurs again.
Obviously, $\lambda=0$ corresponds to the unconstrained random walk, and $\lambda=1$ recovers the results for the strictly SSR 
processes without noise. Note that for $0\leq\lambda<1$, $\Phi_\lambda$ may visit a given site $i$
more than once. This implies in general that the visiting probability $P^{(\lambda)}_{N}(i)$ and the occupation probability $p^{(\lambda)}_{N}(i)$ 
no longer need to be proportional to each other. For that reason we now explicitly compute the occupation probability 
$p^{(\lambda)}_{N}(i)$ for SSR processes with a given noise level. For notation we now suppress $N$ and write $p^{(\lambda)}(i)$.  

Note that $\phi$ produces one realization of $2^{N-1}$ possible sequences of sites $i=1,\cdots,N$, and then stops. The maximum length 
of such a sequence is $N$, the average sequence length is $\l \sim \log N$. In contrast, the unconstrained random walk 
$\phi_R$ has no stopping criterion. 
To avoid problems with mixing processes with different lengths we replace $\phi$ with a process $\phi^{\infty}$ that is identical to $\phi$, 
except for the case when site $i=1$ is reached. In that case $\phi^{\infty}$ does not stop\footnote{In the numerical simulations we stop the process after $M$ re-starts}
 but continues with tossing the $N$ faced dice and thus re-starts the process $\phi$. 
For $\phi^{\infty}$ site $i=1$ becomes both the {\em starting} point of a new single-sequence process $\phi$, and the {\em end} 
point of the previous one (see also Fig. 5 (a)).   
Replacing $\phi$ by $\phi^{\infty}$ in Eq. (\ref{vieh}) ensures that we have an infinitely long, noisy sequence, 
which  is denoted by $\Phi^{(\lambda)}_{\infty}=\lambda \phi^{\infty}+(1-\lambda)\phi_R$.
Successive re-starting gives us the possibility to treat SSR processes as stationary, for which the consistency equation 
\begin{equation}
  p^{(\lambda)} (i)=\sum_{j=1}^{N}  \, {P}(i|j) \, p^{(\lambda)} (j)\quad,
  \label{cond1}
\end{equation}
holds. Here $P(i|j)$ is the conditional probability that site $i$ is reached from site $j$ in the next time step in an infinite and noisy SSR process. 
It reads 
\begin{equation}
	{P}(i|j)=\left\{
	\begin{array}{ll}
	\frac{\lambda}{j-1}+\frac{1-\lambda}{N}	\quad & {\rm for} \  i < j\\
	\frac{1-\lambda}{N}					\qquad & {\rm for} \ i \geq j > 1 \\
	\frac{1}{N}									\qquad & {\rm for} \ i \geq j = 1 \quad . \\
	\end{array}
	\right. 
\label{cond}
\end{equation}
The first line in the above equation accounts for the strictly sample space reducing process, the second line for the unconstrained random walk component, and 
the third line takes care of the re-starting once site $i=1$ is reached.
From Eqs. (\ref{cond1}) and (\ref{cond}) we get 
\begin{equation}
	p^{(\lambda)} (i)=\frac{1-\lambda}{N}+\frac{1}{N}p^{(\lambda)} (1)+\sum_{j=i+1}^{N}   \frac{\lambda}{j-1}  \, p^{(\lambda)} (j) \quad .
\label{eq:P(m)Sum}
\end{equation}
Clearly, the recursive relation $p^{(\lambda)} (i+1)-p^{(\lambda)} (i)=-\lambda\frac1i p^{(\lambda)} (i+1)$ holds, 
from which one obtains 
\begin{equation}
\begin{array}{lcl}
	\frac{p^{(\lambda)} (i)}{p^{(\lambda)} (1)}&=&\prod_{j=1}^{i-1}\left(1+\frac{\lambda}{j}\right)^{-1} 
	= \exp\left[-\sum_{j=1}^{i-1}\log\left(1+\frac{\lambda}{j}\right)\right]\nonumber \\
	&\sim& \exp\left(-\sum_{j=1}^{i-1}\frac{\lambda}{j}\right) 
	\sim \exp\left(-\lambda \log(i)\right) =   i^{-\lambda} \quad .  \nonumber \\ 
	\end{array}
%\label{eq:P(m)recursive}
\end{equation}
$p^{(\lambda)} (1)$ is given by the normalisation condition $\sum_i p^{(\lambda)} (i)=1$, and  
we arrive at the remarkable result,
\begin{equation}
 p^{(\lambda)} (i)\propto i^{-\lambda}.
  \label{eq:P(m)recursive}
 \end{equation}
Note that  $\lambda$ is nothing but the mixing parameter for the noise component. 
For $\lambda=1$ one recovers Zipf's law, $p^{(\lambda=1)} (i)\propto i^{-1}$;  for $\lambda=0$, 
the uniform distribution $p^{(\lambda=0)} (i)=1/N$ is obtained.
For intermediate $0<\lambda<1$ one observes an asymptotically {\em exact} power-law 
with exponent $\lambda$. 
Note that Eq. (\ref{eq:P(m)recursive}) is a statement about the {\em rank distribution} of the system. 
Often statistical features of systems are presented as frequency distributions, i.e. the probability that a 
given site (state) is visited $k$ times, $\tilde{p}(k)$, and not as rank distributions. These are related however. 
It is well known that, if the rank distribution $p$ is a power-law with exponent $\lambda$, 
$\tilde{p}$  is also a power-law with the exponent $\alpha=\frac{1+\lambda}{\lambda}$, see e.g. \cite{Newman:2005}. 
The result of Eq. (\ref{eq:P(m)recursive}) implies that we are able to understand 
a remarkable range of exponents in frequency distributions, $\alpha \in [2,\infty)$, by noisy SSR processes.
Many observed systems in nature display 
frequency distributions with exponents between $2$ and $3$, which in our framework, relates to a mixing ratio of $\lambda>0.5$.
We find perfect agreement of the result of Eq. (\ref{eq:P(m)recursive}) and numerical simulations, Fig. 3 (a). 
The slope of the measured rank distributions in log-scale,  $\lambda^{\rm sim}$, perfectly agree with the 
theoretical prediction $\lambda$. Fitting was carried out by with a maximum likelihood estimator as proposed in \cite{Clauset:2007}.

{\em Convergence speed of SSR distributions.}
From a practical side the question arises of how fast SSR processes converge to the limiting 
occupation distribution given by Eq. (\ref{eq:P(m)recursive}). In other words, what is the distance between 
the sample distribution after $T$ individual jumps in the process $\Phi^{(\lambda)}_{\infty}$  and  $p^{(\lambda)}$, as a function of $T$? 
In Fig. 3 (b) we show the Euclidean distance of the distribution after $T$ jumps $p^{(\lambda)}_{T}$, and 
$p^{(\lambda)}$,   $\left | \left|p^{(\lambda)}_{T}-p^{(\lambda)} \right |\right |_2=\sqrt{\sum_{i\in\Omega}\left[ p^{(\lambda)}_{T}(i)-p^{(\lambda)}(i) \right]^2}$. 
We find that the distance decays as 
\begin{equation}
	\left | \left|p^{(\lambda)}_{T}-p^{(\lambda)} \right |\right |_2 \sim  T^{-\frac12}. 
\end{equation}
The result does not depend on the value of $\lambda$ (see caption). For the pure random case $\lambda=0$, 
our result for the convergence rate is well known and is in full accordance with  the Berry-Esseen theorem \cite{Feller:1966}, which    
accounts for the rate of converge of the central limit theorem for iid processes. 
The fact that for $\lambda=1$ we see practically the same convergence behaviour means that SSR process 
converge equally fast to their underlying limiting power-law distribution. \\

%%%%%%%%%%%%%%%%%%%%%
\begin{figure}[t]
\begin{center}
	\includegraphics[width= 8.5 cm]{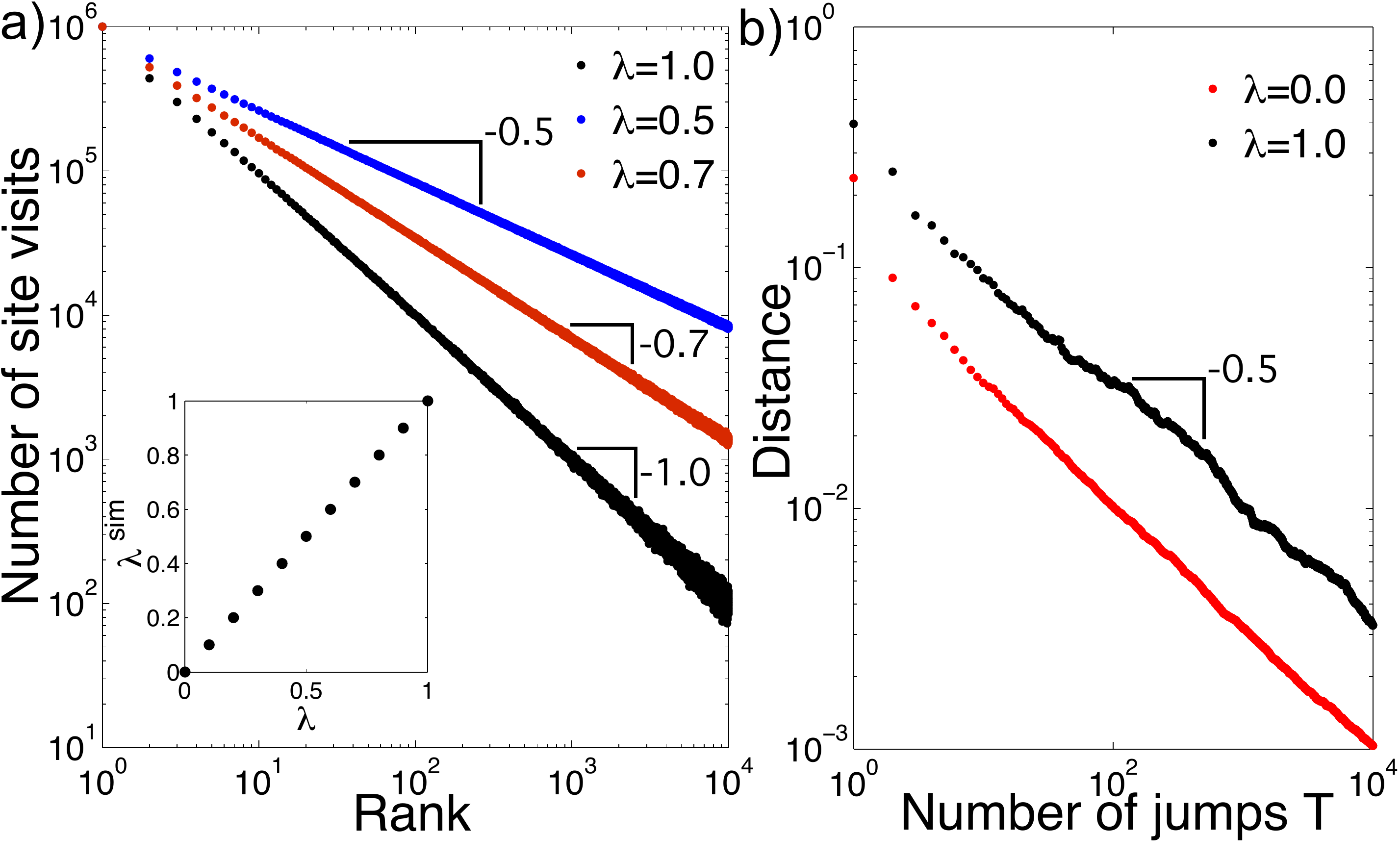}
\caption{(a) Rank distributions of SSR processes with iid noise contributions from simulations of $\Phi^{(\lambda)}_{\infty}$,  for three values of $\lambda =1,0.7$ and $0.5$ (black, red and blue, respectively). Fits to the distributions (obtained with a maximum likelihood estimator \cite{Clauset:2007}) yield $\lambda^{\rm fit}=0.999$, $\lambda^{\rm fit}=0.699$, 
and $\lambda^{\rm fit}=0.499$, respectively. Clearly, an almost exact match with the expected power-law exponents is realised.  
The inset shows the dependence of the measured exponent 
$\lambda^{\rm sim}$  from the simulations (slope), on various noise levels $\lambda$. 
The exponent $\lambda^{\rm sim}$ is practically identical to $\lambda$. 
$N=10,000$, numerical simulations were stopped after $M=10^6$ re-starts of the process.
(b) Convergence rate. The distance ($2$-norm) between the simulated occupation probability (normalised histogram) 
after $T$ jumps in the $\Phi^{(\lambda)}_{\infty}$ process, and the predicted power-law of Eq. (\ref{eq:P(m)recursive}), is shown for $\lambda=1$ (black), and 
the pure random case, $\lambda=0$ (red). Both distances show a power-law convergence $\sim T^{-\beta}$. 
MLE fits yield $\beta=0.512$  and $0.463$,  for $\lambda=0$ and $1$, respectively.  This means that 
both cases are compatible with $\beta \sim 1/2$, and that SSR processes converge equally fast toward their limiting distributions as pure random walks. 
}
\end{center}
\label{Fig:Phi}
\end{figure}
%%%%%%%%%%%%%%%%%%%%%%

{\bf Examples}\\

{\em Sentence formation and Zipf's law.} 
One example for a SSR process of the presented type is the process by which sentences are formed. 
During the creation of a sentence, grammatical and contextual constraints have the effect of a reducing 
sample-space -- i.e. the space  (vocabulary) from which a successive word in a sentence can be sampled. Clearly, 
the process of sentence formation is not expected to be strictly sample-space reducing, and we expect 
deviations from an exact Zipf's law in the rank distribution of words in texts. 
In Fig. 4 we show the empirical distribution of word frequencies of Darwin's  {\em The Origin of Species}, 
which has a first power-law regime with rank exponent  of $\gamma \sim 0.9$. In our framework of the mixed process $\Phi^{(\lambda)}_{\infty}$ 
this corresponds to a mixing parameter $\lambda = 0.9$, indicating that in the process of sentence formation,  
nesting is not perfect, and many instances occur where sample-space can expand from one word to another. 
Note that here $M$ corresponds to the number of sentences in a text. In the simulation we use $N=5,000$ words  
and  $M=10,000$ re-starts. For a more detailed model of sentence formation and SSR processes, see \cite{zipf_books}. 

%%%%%%%%%%%%%%%OriginSpecies
\begin{figure}[t]
\begin{center}
	\includegraphics[width= 7.0 cm]{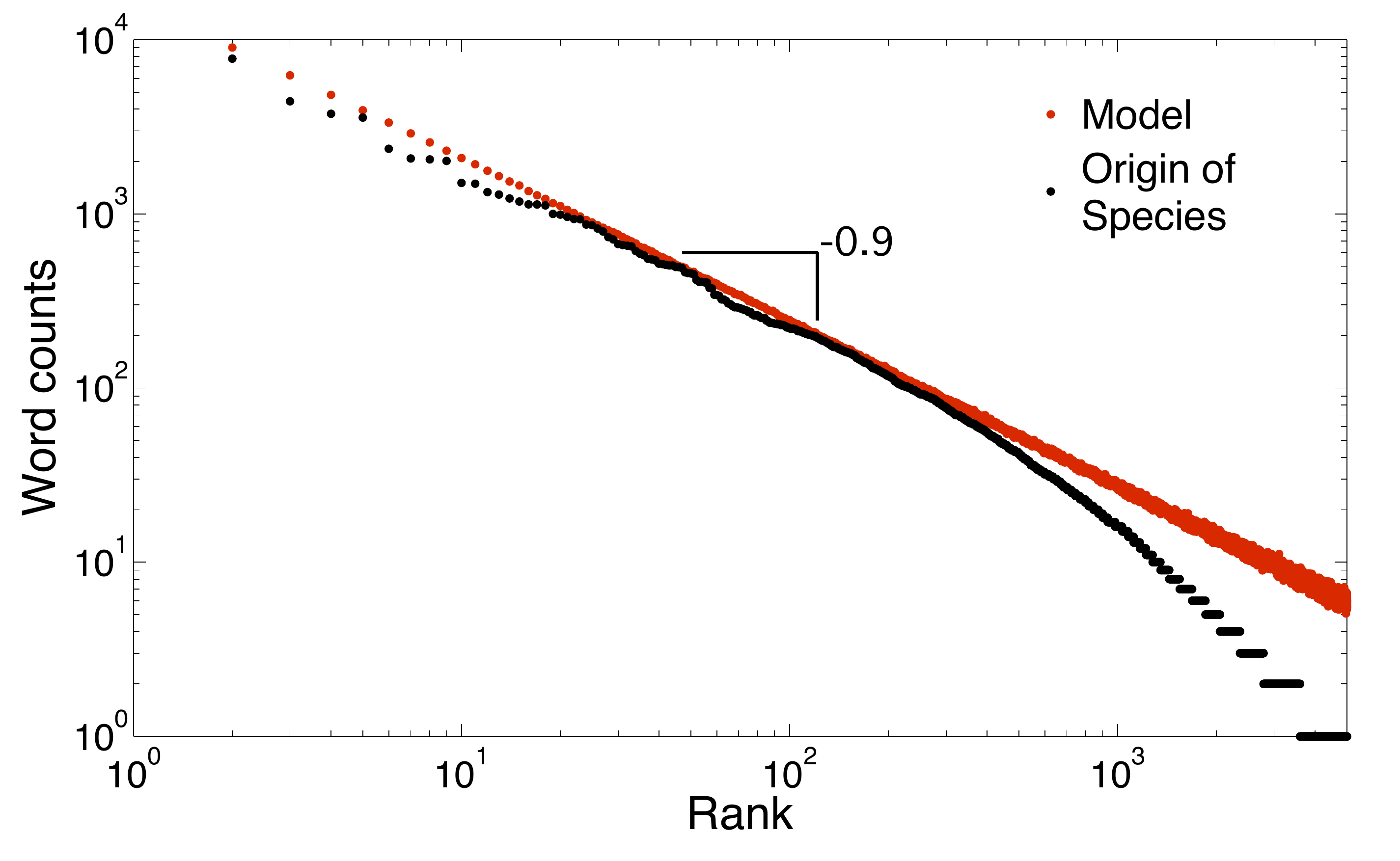}
	\caption{ Empirical rank distribution of word frequencies in {\em The Origin of Species} (black). For the most frequent words the distribution is approximately power-law with an exponent $\gamma \sim 0.9$. The corresponding distribution for the $\Phi^{(\lambda)}$ process with $\lambda =0.9$ (red), suggests a slight deviation from perfect nesting. This means that in sentence formation, about 90\% of consecutive word pairs, sample-space is strictly reducing. Simulation: $N=5,000$ (words), and $M=10,000$ re-starts (sentences).
}
\end{center}
\label{Fig:Phi}
\end{figure}
%%%%%%%%%%%%%%%%OriginSpecies

{\em SSR processes and random walks on networks.}
SSR processes can be related to random walks on directed networks, as depicted in Fig.  5 (a).  
There we start the process from a start-node, from which we can reach any of the $N$ nodes with probability $1/N$. 
At whatever node we end up, we can successively reach nodes with a lower node-numbers until we reach node number 1. 
There, with probability  $p_{\rm exit}$ we jump to a stop-node which ends the process. 
Note that if $p_{\rm exit}=1$, the process runs through one single path and then stops. 
The process is acyclic and finite, there are $2^{N-1}$ possible paths. This network diffusion is equivalent to the 
process $\phi$ above. On the other hand if $p_{\rm exit}=0$, the process becomes cyclic and 
infinite, and corresponds exactly to $\phi^{\infty}$. 
For any $p_{\rm exit}>0$ we have a mixing of the two processes, 
$\Phi^{\rm mix}=p_{\rm exit}\phi+(1-p_{\rm exit})\phi^{\infty}$, which is again cyclic, and the number of possible paths 
is infinite.  
In Fig. 5 (b) we show the result for the node occupation distribution for the process $\Phi^{\rm mix}$ for 
$p_{\rm exit}=1$ (dashed black line) and $p_{\rm end}=0.3$ (solid red line).
The figure is produced from $5\cdot 10^5$ independently sampled sequences generated by $\Phi^{\rm mix}$. 
As expected the distribution follows the exact Zipf's law, irrespective of the value of $p_{\rm exit}$.
The process $\Phi^{\rm mix}$ allows us to study also the rank distribution of paths through the network. 
The path that is most often taken through the network has rank 1, the second most popular path has rank 2, etc. 
Recent theoretical work \cite{Perkins2014} predicts a difference in the corresponding distributions for different values of $p_{\rm exit}$. 
According to  \cite{Perkins2014} acyclic process are expected to show finite path rank distributions of no particular shape. 
This is seen in the inset to Fig. 5 (b) (black dashed line), which shows the observed path rank distribution for the $2^{N-1}=16$ paths. 
For cyclic processes, where at least one node participates in at least two distinct cycles, 
\cite{Perkins2014} predicts power-laws, which we clearly confirm for the cyclic  $\Phi^{\rm mix}$
process with $p_{\rm exit}=0.3$ (red line). 
Note that in our example node $1$ alone is involved in $5$ distinct cycles. 
The process $\Phi^{\rm mix}$ demonstrates the mechanism that produces these power-laws in its simplest form, 
where the probability of long sequences are products of the probability of the finite number of possible sequences which 
they concatenate. 
%Interestingly, this is equivalent to a random language where letters including a stop or whitespace symbol are generated 
%with a certain probability, and words are sequences of symbols between two stop symbols. 
%Random languages of this kind have long been known to produce power-laws \cite{ChomskyMiller}.    

%%%%%%%%%%%%%%%%%%%%%%%%%%%%%%%%%
\begin{figure}[th]
	\begin{center}
		\includegraphics[width=0.95\columnwidth]{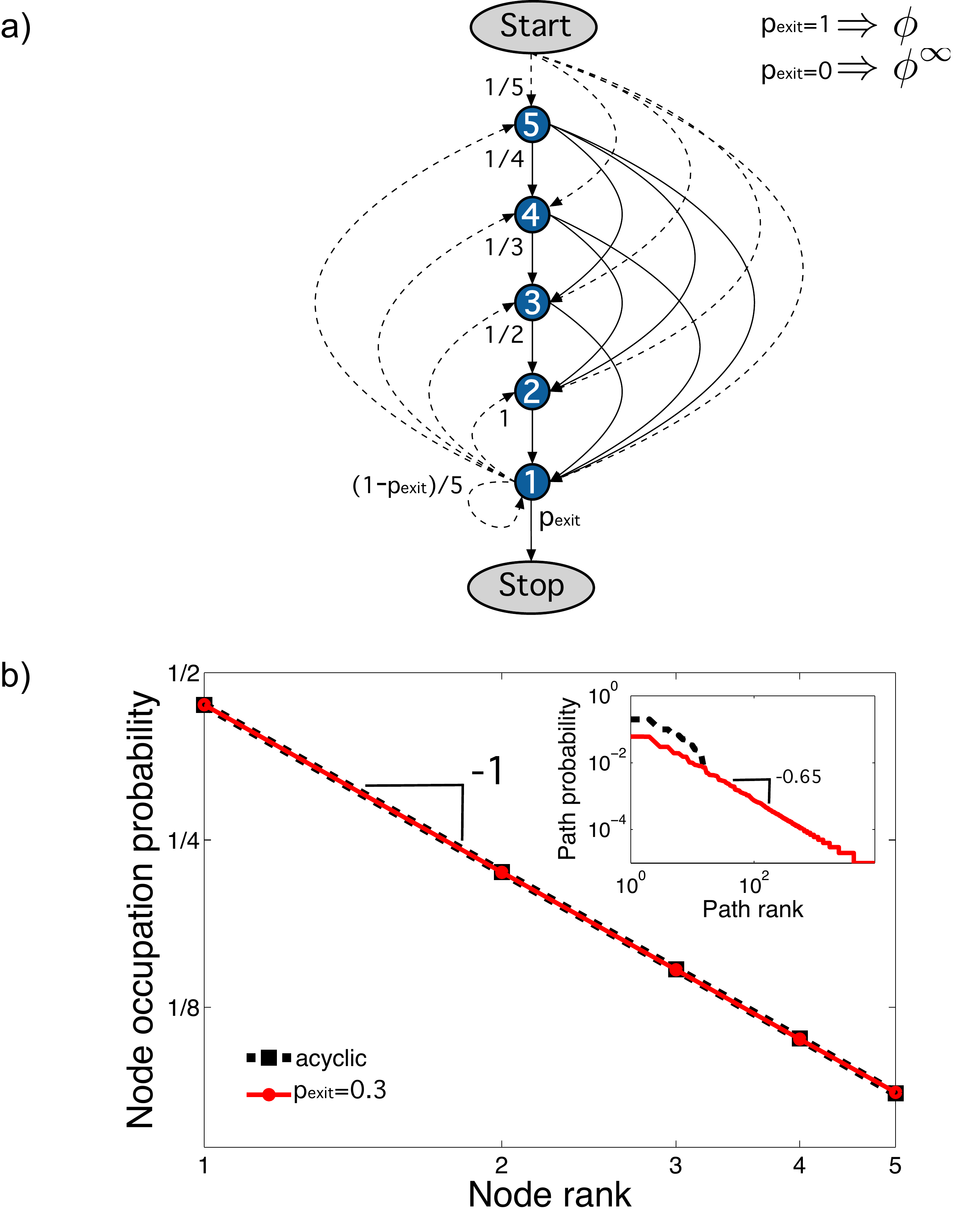}
	\end{center}
	\caption{
(a) SSR processes seen as random walks on networks. A random walker starts at the start node and diffuses through the 
directed network. Depending on the value of $p_{\rm exit}$, two possible types of walks are possible. For $p_{\rm exit}=1$, 
the finite ($2^{N-1}=16$ possible paths) and acyclic process $\phi$ is recovered that stops after a single path; 
for $p_{\rm exit}=0$, we have the infinite and cyclical process, $\phi^{\infty}$. For $p_{\rm exit}>0$ we have the mixed process, 
$\Phi^{\rm mix} =p_{\rm exit} \phi+(1-p_{\rm exit})\phi^{\infty}$.
(b) The occupation probability for $\Phi^{\rm mix}$ is unaffected by the value of $p_{\rm exit}$. 
The repeated $\phi$ (dashed black line), and the mixed process with $p_{\rm exit}=0.3$ (solid red line)
have exactly the same occupation probability $p_N(i)$, 
which corresponds to the stationary visiting distribution of nodes in the $\phi^{\infty}$ network by random walkers. 
(Inset) Rank distribution of paths-visit frequencies. Clearly they depend strongly on  
$p_{\rm exit}$. While the acyclic $\phi$ produces a finite distribution, the cyclic one produces a 
power-law, matching the theoretical prediction of \cite{Perkins2014}. 
For the simulation we generated $5\cdot 10^5$ sequence samples and found $32,523$ distinct sequences for $p_{\rm exit}=0.3$. 
}
\label{Fig:merged}
\end{figure}
%%%%%%%%%%%%%%%%%%%%%%%%%%%%%%%%%

{\em SSR processes and fragmentation processes.}
One important class of ageing systems are fragmentation processes, such as 
objects that repeatedly break at random sites into ever smaller pieces, see e.g. \cite{tokeshi,Krapivsky1994}. 
A simple example demonstrates how fragmentation processes are related to SSR processes.
Consider a stick of a certain initial length $L$, such as a spaghetti, and mark some point on the stick. 
Now take the stick and break it at a random position. Select the fragment that contains the mark and 
record its length. Then break this fragment again at a random position, take the fragment 
containing the mark and again, record its length. One repeats the process until the fragment holding the mark
reaches a minimal length, say the diameter of an atom, and the fragmentation process stops. 
The process is clearly of SSR type since fragments are always shorter than the fragment they come from. 
In particular, if the mark has been chosen on one of the endpoints of the initial spaghetti, 
then the consecutive fragmentation of the marked fragment is obviously a continuous version of 
the SSR process $\phi$ discussed above. Note that even though the  length sequence 
of a {\em single} marked fragment is a SSR process, the size evolution of {\em all} fragments 
is more complicated, since fragment lengths are not independent from each other:  
Every spaghetti fragment of length $x$  splits into two fragments of respective lengths, $y<x$ and $x-y$.
The evolution of the distribution of all fragment sizes was analyzed  in \cite{Krapivsky1994}.
Note that in the one-dimensional SSR processes introduced here we see no signs of multi-scaling. 
However, this possibility might exist for continuous or higher dimensional versions of SSR processes. 

\section{Discussion}

The main result of Eq. (\ref{eq:P(m)recursive}) is remarkable in so far as it explains the 
emergence of scaling in an extremely simple und hitherto unnoticed way. In SSR processes, 
Zipf's law emerges as a simple consequence of breaking a directional symmetry in stochastic processes, 
or, equivalently, by a nestedness property of the sample-space. 
More general power exponents are simply obtained by the addition of iid random fluctuations to the process. 
The relation of exponents and the noise level is strikingly simple and gives the exponent a clear interpretation in terms of 
the extent of violation of the nestedness property in strictly SSR processes. We demonstrate that SSR processes 
converge equally fast toward their power-law limiting distribution, as uncorrelated random walks do. 

We presented several examples for SSR processes. 
The emergence of scaling through SSR processes can be used straight forwardly 
to understand Zipf's law in word frequencies. An empirical quantification of the degree of nestedness 
in sentence formation in a number of books allows us to understand the variations of the scaling 
exponents between the individual books \cite{zipf_books}.  
SSR processes can be related to diffusion processes on directed networks. For a specific example 
we demonstrated that the visiting times of nodes follow a Zipf's law, and could further reproduce very general recent findings of path-visit 
distributions in random walks on networks \cite{Perkins2014}. Here we presented results for a completed directed  graph, however we 
conjecture that SSR processes on networks and the associated Zipf's law of node-visiting distributions are tightly related and 
are valid for much more general directed networks.
We demonstrated how SSR processes can be related to fragmentation processes, which are examples of ageing processes. 
We note that SSR processes and nesting are deeply connected to phase-space collapse
in statistical physics \cite{Hanel:epl11a,Hanel:epl11b,Hanel:2014,Hanel:entropy2013}, where the number of configurations does 
not grow exponentially with system size (as in Markovian and ergodic systems), but grows sub-exponentially. Sub-exponential 
growth can be shown to hold for the phase-space growth of the SSR sequences introduced here.  
In conclusion we believe that SSR processes provide a new alternative view on the emergence of scaling in many natural, social, and  
man-made systems. It is a self-contained, independent alternative to multiplicative, preferential, self-organised criticality, and other mechanisms
that have been proposed to understand the origin of power-laws in nature \cite{Mitzenmacher}. 

{\bf Acknowledgements}.

We greatly acknowledge the contribution of two anonymous referees who made us aware of the relation to network diffusion and fragmentation. We also acknowledge \'Alvaro Corral for his helpful comments.
This work has been supported by the Austrian Science Fund FWF under KPP23378FW.

\end{document}